\documentclass[sigconf,nonacm]{acmart}

\pdfoutput=1
\acmConference[SIGSPATIAL '21]{29th International Conference on	Advances in Geographic Information Systems}{November 02--05, 2021}{Beijing, China}
\acmPrice{15.00}
\acmISBN{978-1-4503-XXXX-X/18/06}

\setcopyright{none} 
\renewcommand\footnotetextcopyrightpermission[1]{} % removes footnote with conference information in first column
\usepackage{graphicx}
\usepackage{hyperref}       % hyperlinks

\usepackage[title]{appendix}

\title{Trinity: A No-Code AI platform for complex spatial datasets}
\begin{document}

\author{C. V. Krishnakumar Iyer, Feili Hou, Henry Wang, Yonghong Wang, Kay Oh, Swetava Ganguli, Vipul Pandey}
\email{{cvkrishnakumar, feilihou, henry_f_wang, yonghong_wang, kay_oh, swetava, vipul}@apple.com}

\affiliation{%
	\institution{Apple}
	\country{USA}
}

%\author{C. V. Krishnakumar Iyer, Feili Hou, Henry Wang, Yonghong Wang, Kay Oh, Swetava Ganguli, Vipul Pandey}
%\email{{cvkrishnakumar, feilihou, henry_f_wang, yonghong_wang, kay_oh, swetava, vipul}@apple.com}

%
%\author{Feili Hou}
%%\affiliation{
%%	\email{feilihou@apple.com}
%%	\institution{Apple}
%%	\country{USA}
%%}
%
%\author{Henry Wang}
%%\affiliation{%
%%	\email{\verb|henry_f_wang@apple.com|}
%%	\institution{Apple}
%%	\country{USA}
%%}
%%%
%\author{Yonghong Wang}
%%\affiliation{
%%	\email{\verb|yonghong_wang@apple.com|}
%%	\institution{Apple}
%%	\country{USA}
%%}
%
%\author{Kay Oh}
%%\affiliation{%
%%	\email{\verb|kay_oh@apple.com|}
%%	\institution{Apple}
%%	\country{USA}
%%}
%%
%\author{Swetava Ganguli}
%%\affiliation{%
%%	\email{\verb|swetava@apple.com|}
%%	\institution{Apple}
%%	\country{USA}
%%}
%
%\author{Vipul Pandey}
%%\affiliation{
%%	\email{\verb|vipul@apple.com|}
%%	\institution{Apple}
%%	\country{USA}
%%}

\begin{abstract}

We present a no-code Artificial Intelligence (AI) platform called Trinity with the main design goal of enabling both machine learning researchers and non-technical geospatial domain experts to experiment with domain-specific signals and datasets for solving a variety of complex problems on their own.  This versatility to solve diverse problems is achieved by transforming complex Spatio-temporal datasets to make them consumable by standard deep learning models, in this case, Convolutional Neural Networks (CNNs), and giving the ability to formulate disparate problems in a standard way, eg. semantic segmentation. With an intuitive user interface, a feature store that hosts derivatives of complex feature engineering, a deep learning kernel, and a scalable data processing mechanism, Trinity provides a powerful platform for domain experts to share the stage with scientists and engineers in solving business-critical problems. It enables quick prototyping, rapid experimentation and reduces the time to production by standardizing model building and deployment. In this paper, we present our motivation behind Trinity and its design along with showcasing sample applications to motivate the idea of lowering the bar to using AI.
 
\end{abstract}

\maketitle

%\section{Introduction and Motivation}

\section{Introduction}
Artificial Intelligence (AI) based solutions have been permeating different fields in recent years due to the explosion of data and rapid strides made in computing power. Almost every field stands to benefit from this advancement provided suitable offerings of AI are identified and its deployment goes both wide and deep. \textit{Wide} to enable more people, ranging from seasoned data scientists to a variety of non-technical domain experts, to directly utilize its power and \textit{deep} to solve intricate and complex domain-specific problems with deeper penetration of AI techniques. In this paper, we introduce a self-service deep learning platform that uses Convolutional Neural Networks (CNNs) based semantic segmentation to go deep in solving complex problems in the geospatial domain. It also offers a code-free environment to enable a wide range of non-technical domain experts and data scientists to perform rapid experimentation and build models without having any specific knowledge of frameworks like Tensorflow \cite{tensorflow}, Python, neural net architectures, or training/prediction infrastructure. It creates a shared vocabulary leading to better collaboration among domain experts, machine learning researchers, data scientists, and engineers. Currently, the focus is on semantic segmentation but it is easily extendable to other techniques such as classification, regression, and instance segmentation.

\subsection{Background}

\subsubsection{Data}

With the increase in the number of smart devices in recent years, a high volume of data containing geo-referenced information is generated and captured. Geo-referenced data is information containing locations on or near the surface of the Earth ~\cite{gis}. Examples of such data include GPS trajectories, remote sensing satellite images, location-based social media, spatial footprints of buildings, roads, elevation data, land cover data, multispectral imagery, and so on. These geo-referenced datasets are valuable inputs for many models to perform tasks such as automatic mapping, urban planning, remote sensing, and land-use detection.

\subsubsection{Problems}
\emph{Geo-feature detection} is one of the tasks in geospatial domain . We define a \emph{geo-feature} as an entity or attribute that has a geo-reference. Examples include entities like buildings, road segments as well as their attributes like \emph{road-type}, \emph{road-segment-centrality}, \emph{road-attributes like one-ways or turn restrictions}, residential regions, \emph{stop signs} etc. Issues could range from missing or misplaced geo-features to incorrectly coded or missing attributes. Some problems are relatively easy to solve while others need a lot of disparate signals from raw data to be consumed and evaluated from different perspectives. 
\subsubsection{Traditional solutions}

Map building processes have traditionally relied on manual labor with tools built to support human editors. While most of these tasks are tedious, some are beyond human reach, given our limitations to fuse multiple modalities of data to make accurate inferences. However, with recent developments in the fields of supervised and unsupervised machine learning, there has been an increase in their application in the field of GIS. There are several studies in the literature that leverage machine learning for tasks in the geospatial domain. Examples include hyperspectral image analysis ~\cite{spectral}, high resolution satellite image interpretation ~\cite{cnn_remotesensing}. Although, deployment of such solutions in the geospatial domain is still limited due to the challenges discussed below.

\subsection{Current Challenges}

\subsubsection{High barrier of entry for Domain experts}
Domain experts are best suited to solve geospatial problems on their own because they have the best intuition of the domain and a deep understanding of the data and derived signals. But processing a high volume of Spatio-temporal datasets and applying Machine Learning (ML) solutions requires specialized skills and therefore has a high barrier of entry rendering non-technical domain experts unable to solve their problems independently. Consequently, non-technical domain experts rely on scientists and engineers to translate their problem statements.

\subsubsection{Non standardized solutions}
Since the problems are seemingly different on the surface e.g separating residential from commercial areas seems vastly different from detecting one-way roads or restricted U-Turn, scientists end up picking locally suitable models or solutions from their repertoire to solve individual problems. Notably, every new experiment requires a distinct set of steps to be performed that includes custom data handling and noise removal, feature extraction, model training, validation, inference, visualization of results, and model deployment. Such ad-hoc handling of data also gives rise to non-standard pre-processing, post-processing, model deployment, and maintenance workflows.

\subsubsection{Bottlenecks and inefficiencies} 
Traditional approaches requiring back and forth communication between domain experts and data scientists are also likely to be bottle-necked by the bandwidth of data scientists and engineers since there is a dependency on the engineers to process data and the scientists to run experiments for different problems. This approach also tends to hamper the ability to collaborate across disparate problems and build shared workflow with the active and hands-on involvement of the domain experts.

\subsection{Our contribution}

Trinity is built to tackle the aforementioned challenges from three angles. Firstly, by bringing information in disparate Spatio-temporal datasets to a standard format by applying complex data transformations upstream. Second, by standardizing the technique of solving disparate-looking problems to avoid heterogeneous solutions and finally, by providing an easy-to-use code-free environment for rapid experimentation thereby lowering the bar for entry. 

\subsubsection{Complex Feature Processing} 

Deep Learning has made significant advances in the area of Computer Vision. Trinity leverages the advancements made specifically in the area of semantic segmentation to solve problems using spatiotemporal datasets. Semantic segmentation ~\cite{semanticsegmentation}  describes the process of associating each pixel of an image with a class label. It has found applications in image processing ~\cite{seg_img}, natural langugage ~\cite{seg_language}, object detection ~\cite{seg_img}, scene understanding ~\cite{seg_scene}, automated map building ~\cite{segmentation_in_maps}, medicine, including brain tumor segmentation ~\cite{seg_med1}, instance aware segmentation ~\cite{seg_med2}, skin lesion segmentation ~\cite{seg_med3} and iris segmentation ~\cite{seg_med4} and many more areas.

A fundamental requirement for semantic segmentation is for the data to lend itself directly to computer vision algorithms. There are some datasets in geospatial domains that fall in this category e.g. satellite, aerial, and drive imagery, but the others that are not directly suitable for aforementioned techniques e.g. sequence data that encodes the motion of devices in GPS traces and temporal data sets. 
\subsubsection{Generalization of complex datasets as image-like channels}
One of the key aspects of Trinity is the way it leverages such complex datasets after transforming them into CNN-friendly RGB (Red-Green-Blue)-like channels. Some transformations can be simple aggregation into pixel values e.g. densities of walking and driving GPS traces per pixel of a certain size \cite{probe} but can go all the way to self supervised representations called \textit{embeddings}. These representation are laid out as \textit{embedding fields}. These channels are pre-generated and stored in a channel store and shared across multiple models and different problems. Embedding fields help in retaining relevant information in the data and also bringing it to the fold of computer vision. Pre-generation of channels also takes away the need for repeated processing of the data for different models in various stages of its life.

\subsubsection{Standard Deep Learning Kernel}
Trinity standardizes the way most of the geospatial tasks are solved by formulating them as semantic segmentation problems. This also opens up the avenue of reusing off-the-shelf algorithms and CNN architectures like VGG, UNet, RCNN, ResNet, Mask R-CNN, etc.,  benefiting from any advances made in this area in both industry and academia. Trinity packages a handful of standard CNN-based segmentation architectures in its deep learning kernel for the users to pick from. The user picks suitable channels from the channel store and standard architectures from the kernel to run train their models (called \textit{experiments} in Trinity). Alternatively, they can choose to leverage AutoML functionality (Sec \ref{sec:automl}) to let Trinity pick the best network architectures for their use case. We have successfully formulated previously disparate-looking problems like detecting one-way roads, parking lots, presence of traffic control devices (eg. stop signs and traffic lights), medical centers, golf courses, etc. as semantic segmentation problems and solved them using spatiotemporal channels in the same way.

\subsubsection{Platform}
By wrapping a robust software system around the kernel and channel store we have opened it up for anybody to run their experiments at scale using a variety of channels and CNN architectures without writing a line of code. This includes an intuitive UI for experiment management, scalable data storage (called Trinity channel store), containerization of the kernel for deploying on GPU clusters, and a suite of standard pipelines for data preparation and large-scale predictions on Hadoop using Spark. This facilitates data sharing, model sharing, low time to market, standard ML/DevOps, and seamless collaboration between domain experts and scientists. The biggest benefit is the ability to run quick and parallel experiments with minimal overhead.  We witnessed first-hand and strongly agree with the observation in \cite{fblearnerflow} that the largest improvements in accuracy often are tied to the ability to perform quick and rapid experimentation.

\section{Related Work}

In this section we take a look at products similar to Trinity in some aspects. We group these into different categories.

\textbf{Geospatial Platforms:} Traditionally, geo-feature detection in the geospatial domain usually involved heavy preprocessing like Map-matching \cite{mapmatching} and was typically based on heuristic techniques. In these approaches, typically data was heavily preprocessed, features were extracted and traditional machine learning models were trained to detect the geo-features. However, recently there have been several instances of solutions applying semantic segmentation. For instance, \cite{deeplearning_buildingdetection} focuses on identifying buildings from Satellite imagery and \cite{objdetection_satelliteimagery} describes object detection using multi-scale CNN. However, these solutions are built on a specific type of data modalities like satellite imagery and specific types of tasks like object detection. Moreover, with the standard approaches, fusing image and non-image-based channels can be a challenge. Trinity is built to address these problems by expanding the scope of deep learning algorithms to multiple modalities of data and can be used to address a diverse variety of problems in a streamlined fashion. 

\textbf{Pipeline Frameworks:} There are several general-purpose data pipeline frameworks that help streamline machine learning products. For instance, FB Learner Flow \cite{fblearnerflow} supports experiment management and allows engineers to define \emph{workflows} using decorators. Similarly, the ML Data Platform \cite{mldp} focuses on version management of algorithms, reproducibility of ML experiments, and integration of ML frameworks besides maintenance of data provenance for lineage tracking. However, these platforms aim to help users build and maintain their pipelines. Trinity is designed to work on a different layer of abstraction and aims to also support the non-technical users by providing a zero-coding platform where the pipeline construction and maintenance are abstracted away.

\textbf{ Generic ML Platforms:} Tensorflow Extended \cite{tfx} is a  TensorFlow-based general-purpose machine learning platform that aims to standardize the components, simplify the platform configuration, and reduce the time to production. Likewise Amazon's SageMaker \cite{ sagemaker} and Azure ML Pipelines \cite{azureml} are general-purpose tools for building ML workflows. Similarly, Uber's Michelangelo \cite{michelangelo} focuses on streamlining the entire ML workflow, including management of data and features, shared feature stores, and a workflow system that orchestrates data pipelines in addition to container deployment. Linkedin's ProML \cite{proml}, Netflix's Metaflow \cite{metaflow} and Onepanel \cite{onepanel} are other solutions in the same direction. Google Cloud AI Platform \cite{googleautoml} takes a more expansive approach supporting data preparation, model building, validation, and deployment using the MLOps \cite{mlops} best practices.

Trinity is also designed to help streamline the entire ML workflow but it operates in a focused and limited scope of geospatial applications. It is not a general-purpose ML platform that requires users to code their models. The feature and model management are abstracted away from the users so that the users can focus on building the model using a self-service paradigm. Moreover, sophisticated features (including task-agnostic embeddings) are generated and prepopulated for the use of users.

\textbf{Self Serve domain-specific platforms:} Recently, there have also been several no-code products and frameworks that promote a similar self-service model. Some examples are Google Cloud AutoML \cite{googleautoml}, Landing Lens \cite{landinglens} and Matroid \cite{matroid}. However, these focus on specific verticals, with imagery being the primary channel for Landing Lens, Matroid, and Google Cloud AutoML Vision. To the best of our knowledge, we have not seen an end-to-end, code-free, self-service platform for deep learning designed for complex spatiotemporal datasets in the geospatial domain that extends to both technical and non-technical users alike. This is the gap that Trinity expects to fill.

\section{Platform Overview}

\subsection{Design Principles}
The core requirements underpinning the design of Trinity are described in this section. 

\textbf{Utilize disparate datasets:} Provide the ability to use visual (imagery), structural (graphical and point,) and behavioral datasets \cite{probe}, either individually or in conjunction, for detecting patterns and anomalies. For instance, users must be able to fuse multiple different sources of data such as satellite imagery, location traces, and road network to detect geo-features like Parking lots.

\textbf{Complex problem formulation:} Enable users to formulate complex problems in a manner that brings them into the manifold of a standard set of algorithms. For example, using a common platform, semantic segmentation should be leveraged to detect diverse geo-features ranging from one-way roads to buildings and tennis courts, and all the complexity should be abstracted away from the users.

\textbf{Streamline processes:} Enable a standardized and streamlined set of process workflows that can be applied uniformly for different applications to ensure easier maintenance. 

\textbf{Low barrier to entry:} Enable users with no technical experience to validate their ideas/hypotheses, run experiments, and rapidly test their solutions without depending on data scientists and engineers. This brings the power of deep learning to users most familiar with the domain and the associated datasets by abstracting out the technology.  

\textbf{Extensibility and modularity:} Enable easy extension of the platform so that adding new algorithms and architectures is simple. Facilitate modularization of the platform so that features, models, and problem formulation can evolve independently of each other. 

\textbf{Quick experimentation:} Enable users to quickly test their hypothesis. The idea is to facilitate users get to a minimum viable solution very rapidly rather than spend weeks to arrive at a state-of-the-art solution.

\subsection{Technology Stack}

\begin{figure}[hb]
	\begin{center}
		\includegraphics[scale=0.30]{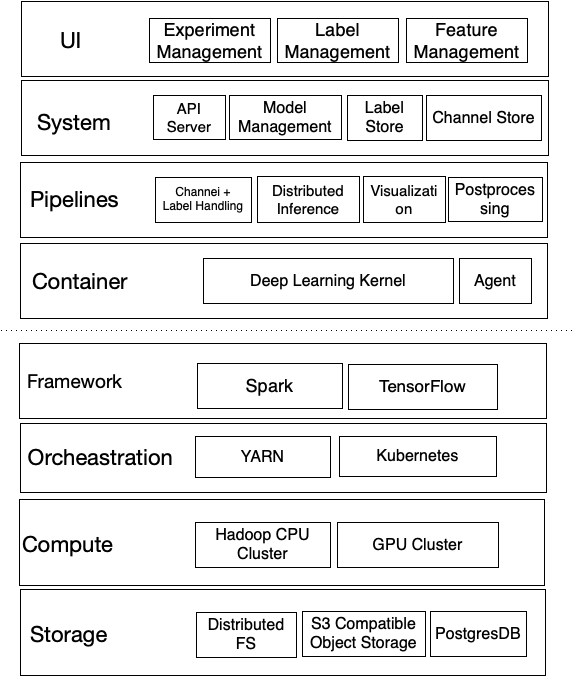}
	\end{center}
	\caption{Trinity Component Layers}
	\label{fig:layer_components}
\end{figure}

Trinity is composed of data pipelines, an experiment management system, a user interface, and a containerized deep learning kernel that we discuss later. This section focuses on the technology stack as depicted in the bottom half of Figure ~\ref{fig:layer_components}. These infrastructure pieces are used as building blocks to the system that is the subject of  Section ~\ref{sec: systemoverview}.

\textbf{Storage:} Trinity is designed to leverage different types of storage implementations. While its feature store is maintained in an S3 (Simple Server Storage) compatible storage, any intermediate data, inputs, and processed predictions are stored in a distributed file system (HDFS). Metadata related to the experiments, including versions of models, are stored in an instance of a PostgreSQL DB  running on an internal cloud infrastructure.

\textbf{Compute:} Internal compute clusters hosting GPU and CPU instances are leveraged for elastic usage.

\textbf{Orchestration:} For portability and packaging, the training is containerized using Docker and orchestrated by Kubernetes running on the GPU Cluster. Large-scale distributed predictions are carried out on CPU clusters orchestrated by YARN.

\textbf{Frameworks: } As of this writing, we use Tensorflow 2.1.0 for training our deep learning models and predicting with them. We leverage Spark on Yarn for data preprocessing, channel processing, label handling, distributed inference, visualization of predictions, and post-processing of results for downstream applications.

\section{Architecture and components}
\label{sec: systemoverview}
\begin{figure}[!hbt]

		\includegraphics[scale=0.24]{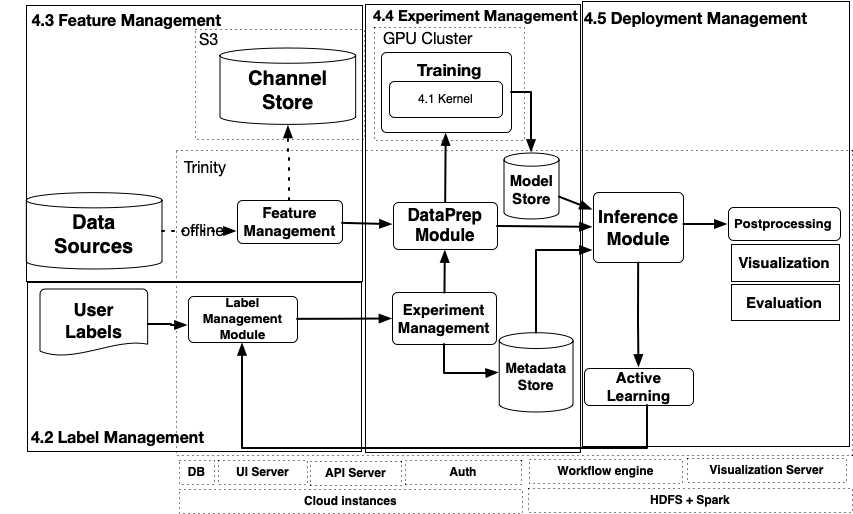}
	\caption{Trinity Modules}
	\label{fig:trinity_components}
	
\end{figure}

Figure~\ref{fig:trinity_components} gives an overview of the platform and interactions among the modules.
This section describes each module in detail.

\subsection{Deep learning kernel}

The deep learning kernel is at the heart of the platform and encapsulates neural net architectures for semantic segmentation and provides for model training, evaluation, handling of metrics, and inference. The kernel is currently implemented in TensorFlow but can easily be swapped for using other frameworks. 

\subsubsection{ Network architectures}

In Trinity, the kernel provides multiple encoder-decoder architectures for segmentation, such as FCNs~\cite{fcn}, SegNet ~\cite{segnet} and UNet ~\cite{unet} that can be selected from a catalog by the users for any given task. This module is designed to be flexible and extensible such that new and promising segmentation models can be plugged in with ease and the kernel can evolve independent of the rest of the platform, and support multiple kinds of tasks.
%\subsection{Deep Learning Kernel}

\subsubsection{Types of tasks}

Trinity is designed to support binary and multi-class semantic segmentation tasks. Binary segmentation has just two classes: foreground representing the object of interest (such as a parking lot) and the background class. It aims to separate the foreground from the background pixels thereby detecting the presence or absence of the target class. It can be used for a variety of tasks including detection of the road centerlines (Figure ~\ref{fig:centerline}) and residential areas (Figure ~\ref{fig:residential}). The multi-class segmentation task gives the ability to detect multiple classes of the target. An example of this is type-of-road detection that extracts different types of roads such as alleys, freeways, and ramps.

The platform also supports multi-task segmentation (also called \emph{multi-label} segmentation ), where each pixel can potentially have multiple labels. In this process, multiple tasks, like detection of pedestrian crosswalks,  walkable roads, and road class (like Freeways, ramps, residential roads) can be trained together by the same neural network. The advantage is that this promotes cross-task learning. For instance, the model can learn the correlation between walkable roads and pedestrian crosswalks. Multi-task models are also useful in other use cases like learning different road casements, buildings, and road center-line detection together. One of the key benefits of doing this is that tasks having a lesser number of labels can benefit from the other tasks. There are also speed benefits since one single inference can generate multiple different features at once thereby saving time. 

Additionally, there is support for \textit{multi-modal learning }for mixing different modalities of data. For instance, users can mix satellite imagery with road network data to more accurately predict buildings, pedestrian crosswalks, and road signage.

\subsubsection{Training and Metrics}

Trinity is a supervised learning platform and therefore the training phase consists of model fitting based on input data and labels. We discuss label and feature management in Section ~\ref{sec:labelmanagement} and Section ~\ref{sec:channelstore} respectively.

In the learning phase, this kernel is packaged in a docker container and the training is launched on a GPU cluster orchestrated by Kubernetes as discussed above. Training data is either locally cached on GPU nodes or streamed directly from HDFS depending upon its size.   

Different metrics on training and validation data set such as per-task accuracy, precision, recall, loss, and fIoUs are logged during the training phase and can be visualized on a dashboard. Additionally, there's support for the warm-start of models for transfer learning. Warm start models are useful especially when the models are ported across geographical markets with significantly different behavior or landscapes. Models are saved every few epochs using the standard \textit{SavedModel} format of Tensorflow.  The kernel has several hyperparameters and the default configuration uses adam optimizer and cross-entropy loss with 30\% hold out validation set, but it is very straightforward to change the optimizer and loss functions or add new ones. Additionally, as described in the Section ~\ref{sec:automl}, Hyperparameter optimization support is also provided to intelligently select the best hyperparameters for the model.

\subsubsection{AutoML: Hyperparameter Optimization}
\label{sec:automl}
\begin{figure}[ht]
	\begin{center}
		\includegraphics[scale=0.24]{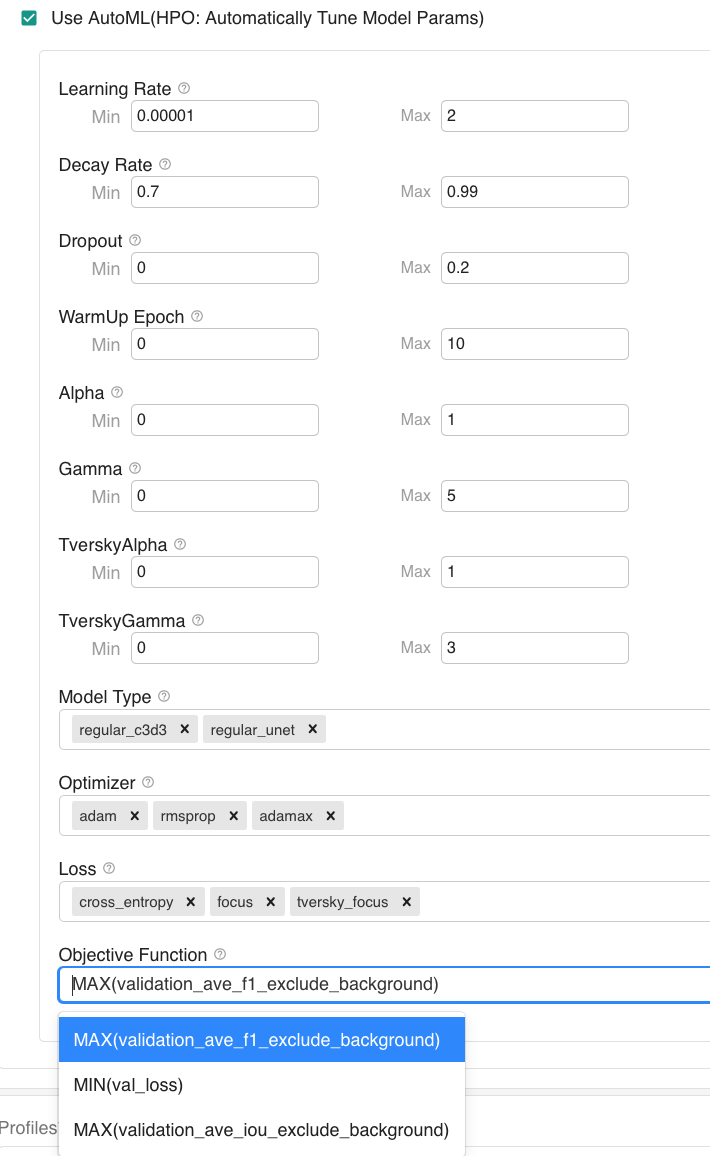}
	\end{center}
	\caption{Hyper Parameter Optimization}
	\label{fig:automl}
\end{figure}

Trinity supports hyperparameter optimization as a first-class feature of the platform. It enables users to enhance their productivity by taking up the automatic intelligent hyperparameter selection. The user provides the ranges of the different hyper parameters as part of the experiment setup as shown in Figure \ref{fig:automl}. Automatically, multiple parallel trials are launched to select the best hyperparameters. The best hyper parameters are then used to generate the best model. All of this is tracked and managed automatically.

\subsubsection{Inference}

The kernel is packaged with all dependencies inside a virtual environment and run on a Spark cluster during inference to leverage data locality and massive CPU-based compute resources.  The kernel is detailed in Section ~\ref{sec:distinference}.

\subsection{Label Management }
\label{sec:labelmanagement}
The label management module is responsible for making problem formulation, in the form of label generation and collection, convenient for the users.  Labels used in Trinity are geometry objects like points, lines, or polygons. A variety of geo-feature detection tasks can be addressed with semantic segmentation if they are appropriately formulated using suitable labels. Some labels like road center-lines or polygons of pedestrian crossings are easy to represent, while others like labels for turn restrictions and one-way roads need customized label representation. The platform provides tools and frameworks developed in-house that enable users to bring their tasks to the domain of semantic segmentation. It also supports multiclass segmentation where the geometries are coupled with class tags. These geometries represent the geo-features of interest from the user perspective. An example is the type of road detection as represented in Figure ~\ref{fig:fow}. 

The label management module converts geometries into raster images. In the raster, a pixel represents a Zoom 24 tile \footnote[1]{\label{f1} A tile resulting from viewing the spherical Mercator projection coordinate system(EPSG:3857) \cite{embed1} of earth as $2^{24}\times2^{24}$ grid. The grid corresponds to a spatial resolution of approximately 2.38 m at the equator.}  and the raster label image of 256 x 256 pixels - same dimensions as any input channels discussed below.

\subsubsection{Types of labels}
\begin{figure}[hb]
	\begin{center}
		\includegraphics[scale=0.25]{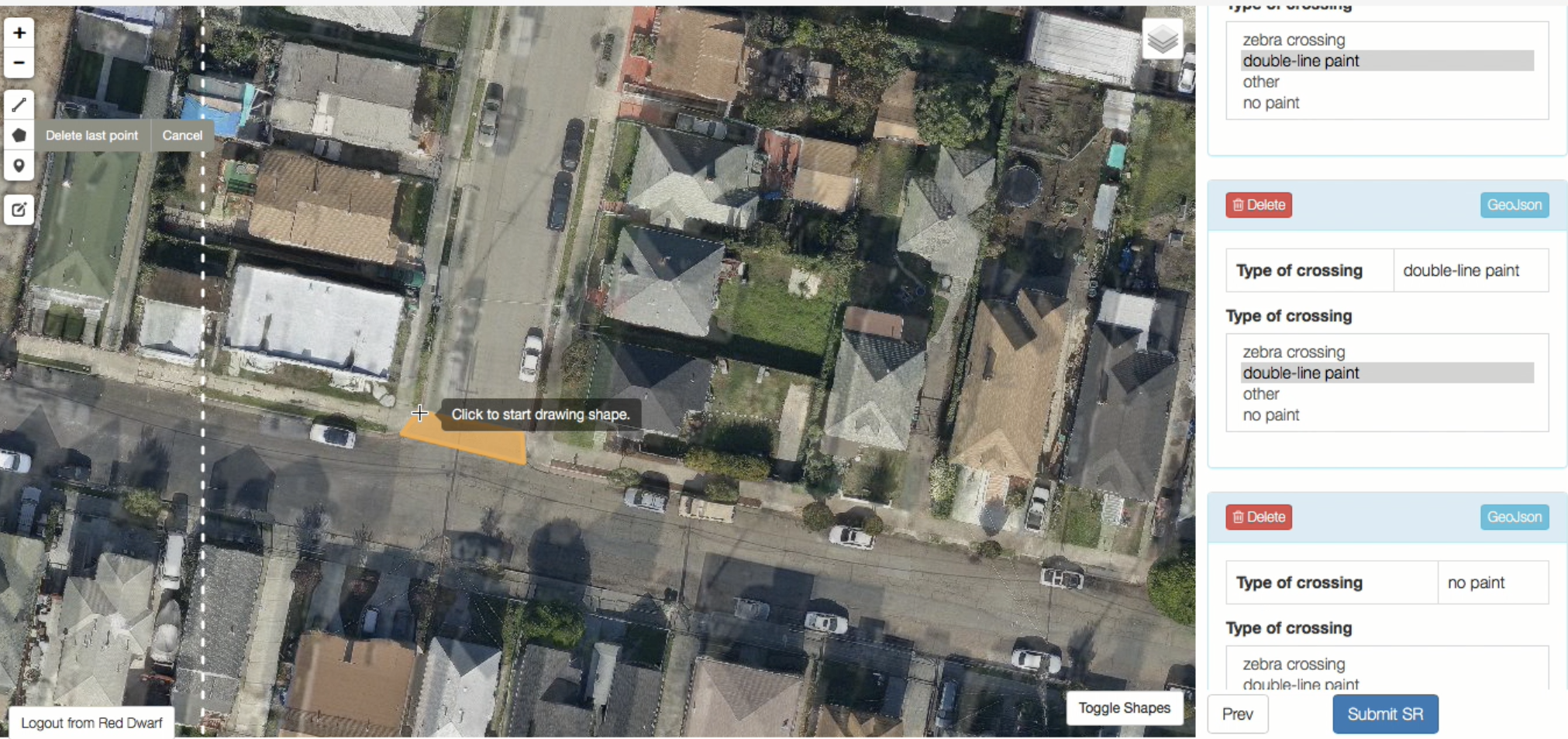}
	\end{center}
	\caption{In-house labeling tool}
	\label{fig:lava}
\end{figure}

Trinity supports different kinds of labels. Firstly, there are \textit{Uploaded Geometries}, where users are able to upload their own geometry files right from the Trinity UI. The labels, represented by geometries (which can be points, lines, or polygons) are ingested in the \emph{Well-Known-Text(WKT)} format. Secondly, Trinity supports hand annotated geometries too. Trinity has tight integration with proprietary in-house tools that can be used for collecting labels. Figure ~\ref{fig:lava} depicts one UI that is used to hand annotate geometries for specific geo-feature detection along with the specification of the class tag that could potentially be used for multi-class segmentation.   

\subsubsection{Active Learning}

Trinity supports \textit{active learning }as a first-class feature of the platform. We see active learning with the human-in-the-loop as an integral part of our project life cycle. Active learning loops enable users to get maximum value from the labeling efforts by judicious selection of labels based on uncertainty involved in the predictions. This paves the way towards faster iterations.

The \textit{active learning module }enables users to use their predictions as starting points to generate new labels. Once the user performs inference on a dataset, the active learning module selects the instances that the current model is not confident on to be labeled. A labeling task is automatically created and shared with the users in the label management tool. Once the requested data is labeled the active learning module creates a new clone of the current experiment with these additional labels. This improves the quality of the model over time as it enables labels to get better with in line feedback from the users

\subsection{Feature Management}
\label{sec:channelstore}

Having discussed deep learning kernel and label management, this section focuses on the representation of the input features.

\subsubsection{Feature Representation}

Input features in Trinity are expressed as an ordered sequence of \emph{Channels}. A \textit{channel } is a two dimensional matrix consisting of \textit{pixels}. The typical size of the channel is $256 \times 256$ pixels where each pixel is represented by a numerical value at the granularity of a zoom 24 $tile^1$.

\begin{figure}[H]
	\begin{center}
		\includegraphics[scale=0.25]{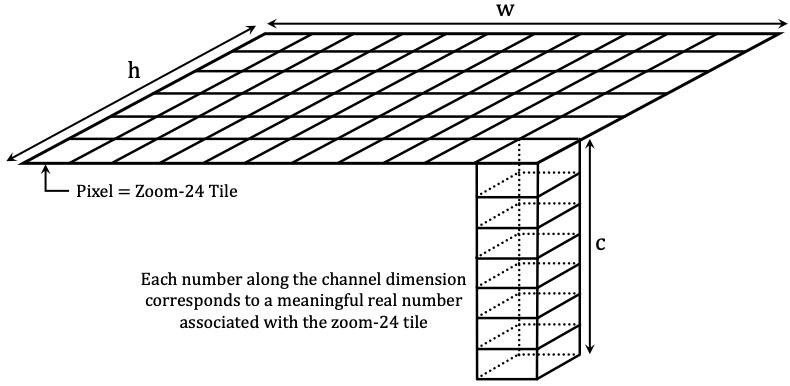}
	\end{center}
	\caption{Profile Representation}
	\label{fig:profile}

\end{figure}

Related channels are saved as an atomic group such that they always appear together in a predetermined order whenever selected. Such grouping of channels is called a \emph{profile} as depicted in Figure \ref{fig:profile}. A profile can have one or more channels.  For example, the satellite image profile contains the three RGB channels. Trinity exposes a catalog of profiles for users to choose from for their experimentation. These profiles are generated from data aggregated across a unit of time in a secure and privacy-preserving fashion. 

\subsubsection{Types of channels}

There are different types of channels that Trinity supports. 

\textbf{Imagery channels:} The image profiles contain 3 channels - one each for red, green, and blue for each pixel. The images are also stored at a Zoom 24 pixel level just like other channels by up or down sampling. This set of layers can be crucial in some applications such as detection of buildings or sports fields.

\textbf{Simple aggregates:} These profiles and their channels are generated from probe data that captures user motion and behavior. Different types of information are extracted from the logs and aggregated per pixel across a unit of time in an anonymous and privacy-preserving fashion. This representation facilitates the conversion of spatiotemporal data to image-like channels that are compatible with CNNs. 

\textbf{Complex aggregates: }\label{complexagg} Information in the data that cannot be summarized as simple aggregates are extracted using complex transformations and persisted as pixel values in the profile store. An example is the representation of time series data of pixel-wise densities across a unit of time as a set of Fourier components to capture seasonality and temporal patterns for each pixel. This profile is helpful in separating out residential areas from commercial ones.

\textbf{Embedding fields: } Certain behavioral information is even more difficult to squeeze out of location-based behavioral data. In such cases,  upstream models are trained to generate \textit{embeddings } per pixel from the raw data. The images that are composed of these per-pixel embeddings and used in Trinity are called ``embedding fields". An example is the \textit{reachability embeddings} that encodes the transition behavior from each zoom-24 tile to all the locations in its neighborhood. Reachability embeddings are self supervised representation for each tile derived from the associated transition probability matrices.  These embeddings are similar to word embeddings used in the context of Natural Language processing. 

\textbf{Transient channels:} Transient channels are one-time use channels, generated by upstream pipelines and injected into the Trinity workflow when needed. Transient channels are not persisted in the channel store but supplied by the production pipeline of a deployed model and are generally based on structural data that changes often, for instance an evolving road network substrate. 

\textbf{Trinity output based channels:} Since there is a consistency between input and output image sizes (barring the number of channels) some Trinity model outputs are fed back into the profile store for other experiments or products to consume. For example, the predicted probability of a region being residential that is output from Trinity is fed back in as a profile and is used by other models to utilize that information.

In addition to the above, Trinity is fully flexible to persist and serve a diverse set of channels including synthetic channels ~\cite{vaeinfogan}. For instance, attributes like elevation, weather patterns, crime data, popularity, places of interest, traffic profiles can all be represented as separate profiles and saved in the store for users to make use of. The profile store is a standalone subsystem that can also be used in task-agnostic analyses of the different regions even while lending itself to non Trinity-based use cases.

\subsubsection{Channel Store}

Channel Store is where \emph{profiles} are saved. It is therefore also called the \emph{profile store}. Channels are stored in a sparse format and the store is implemented on an S3 compatible storage in internal cloud infrastructure. The channel store also decouples the details of the storage of channels from their usage.  Each profile also has associated metadata, that is exposed to the user via a catalog and is used by the users to select the profiles needed for their experiments. The seamless sharing of profiles and channels paves the way for data reuse and helps avoid redundant re-computation and preprocessing of data.

\subsubsection{Benefits of predefined channels}

There are several advantages to having predefined channels.

\textbf{Sharing and reuse:} Channel store is designed to promote collaboration and channel reuse among diverse users.  It enables channels generated for one use case to be used for several others. Channel catalog gives users the freedom to experiment with any number and combinations of profiles they find suitable for their project. For instance, the stop sign detection model (Figure \ref{fig:stopsigns}) uses the same heading channels that are used by the one-way detection model (Figure \ref{fig:oneway}).

\textbf{Decoupling generation and use of channels: } Due to the modular nature of profiles users are spared from processing input data directly. Instead, they can raise their level of abstraction and focus their attention on mixing and matching ready-made channels that can get them to their desired goal. Scientists and engineers who are building or loading these channels can work independently of downstream applications. Likewise, the users who are focused on the application do not need to worry about scalable ways of generating channels.

\textbf{Rapid experimentation:} Predefined channels that are populated asynchronously and are always available with certain guarantees reduce the cost of generation of input data thus speeding up experimentation.  Wherever possible the platform enables the users to aggregate channels across a duration of time on the fly providing the flexibility to select the amount of data they need for their experiments. The same user for the same project may launch different experiments with not just disparate sets of profiles but also with the same set of profiles/channels aggregated for a longer or shorter duration. Rapid experimentation is considered to be a key to finding a stable solution faster.

\subsection{Experiment Management}

Trinity is designed to facilitate efficient experiment management that promotes zero coding and rapid iterations. In order to achieve this, the platform provides a simple and intuitive Web UI that enables the users to develop a self-service model for both training and inference. Every trained model is part of an experiment that is set up and carried out by the user. An experiment comprises of three main attributes based on which a model is trained, viz. labels, neural net architecture, and channels. Out of the three components of an experiment, the neural network architecture and the channels are selected by the user from their respective catalogs that are provided in the UI. Labels are created and managed in the ways described above.

\subsubsection{User Interface}

\begin{figure}[hb]
	\begin{center}
		\includegraphics[scale=0.23]{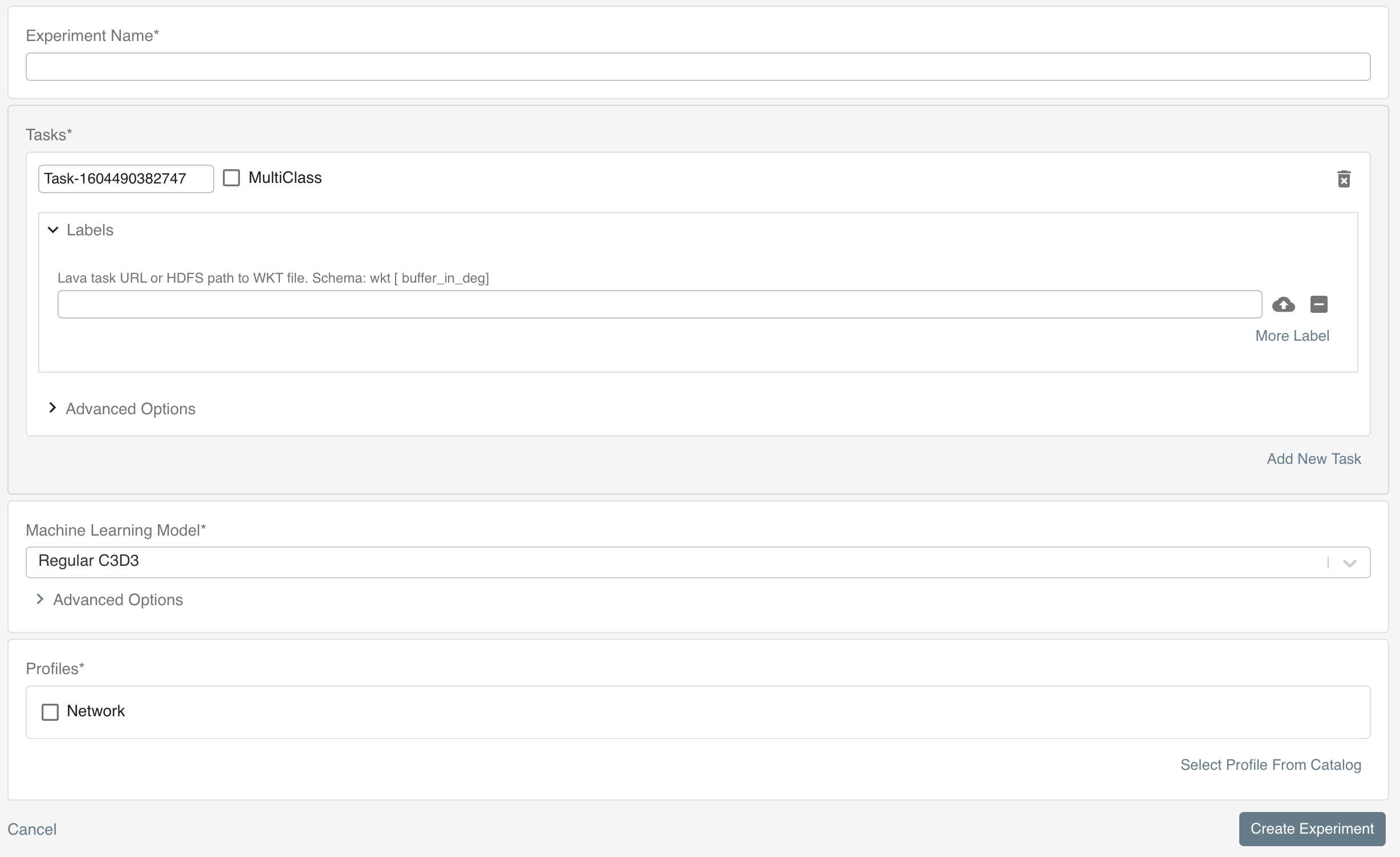}
	\end{center}
	\caption{Screenshot of the Trinity Experiment UI}
	\label{fig:expview}
\end{figure}

The user creates a project to solve a given task of feature detection and has the ability to create and carry out as many experiments as preferred. The \textit{project view }of the UI provides a convenient way for users to search, monitor, and compare the experiments. The \textit{Experiment view} gives users the ability to select profiles from the profile catalog, monitor data preparation and model training,  sample prediction and visualization, validate training and prediction data distributions and view the overall status of the experiments. Users can promote experiments deemed to be promising and can clone any experiment to build on top of others' experiments. Moreover, experiments can be tagged and users can add notes to them for future reference. A screenshot of the Trinity experimentation UI is provided in Figure ~\ref{fig:expview}. Once the training is underway, the user has the ability to pick a stable model and launch test predictions for a target region. The ability to launch predictions using intermediate models helps the user get a sense of how well the model generalizes and makes appropriate changes as required. 

\subsubsection{Data Preparation Module}
The Data Preparation Module consists of multiple pipelines responsible for preparing data for training and inference phases.  These pipelines are implemented in Spark and run using an in-house workflow engine.  They are responsible for tasks such as label extraction and handling, channel extraction, data transformation, and channel-based ``image" creation. The end product of this module is a dataset that can be directly ingested by the kernel for either training or for prediction. 

\subsubsection{API Server}
The experiment management module has an API server that runs on our internal cloud offering and exposes end-points to perform authentication, job submissions for training and predictions, and status checks. The API server is backed by a PostgreSQL database. 

\subsubsection{Metadata Management}
All of the experiment metadata, including the channels used, the network architecture used, predictions, and pointers to the resultant model files are stored in the database for high availability and easy access.

\subsection{Deployment Management}

\subsubsection{Scalable Distributed Inference}
\label{sec:distinference}

The distributed inference is the primary mode in which trained models are used to predict on new and unseen datasets in a scalable fashion. The output of the inference is a confidence heatmap on a per-pixel level per class for each task. Trinity is geared to leverage the natural partitioning and data locality of the distributed file system to package the inference code and ship it to the Spark executors for inference. Tensorflow based prediction code is run in a python virtual environment that is created inside every executor. The inference thereby happens in a scalable data-parallel fashion and the results are stored in the distributed file system.

\subsubsection{Visualization}

Once the inference is complete Trinity automatically generates raster layers, also called \emph{heatmaps}), to visualize the predictions. They help the users to pan and scan to check for the quality if required either before post-processing the output for downstream consumption or for debugging purposes later on. Figure ~\ref{fig:binarysegmentation} shows us the examples of binary segmentation carried out for detecting pedestrian crossings. It is a heatmap of confidences on a per-pixel basis with confidences below a predetermined and suitable threshold filtered out. For multi-class Segmentation, the dominant class for each pixel is displayed in the visualization whereas for multi-task predictions sets of heatmaps for each task are generated. An example of this is the one-way detection model represented in Figure ~\ref{fig:oneway}.

\subsubsection{Post-processing}
\label{postprocessing}

In addition to the prediction, the platform provides workflows for standard post-processing of results. Post-processing is needed to make the results of predictions usable for downstream processing for taking appropriate actions such as identifying missing geo-features, correcting existing issues, etc. either manually or as auto-fixes. 

Users can choose from different strategies for post-processing. The most popular ones include \textit{Vectorization}, which is the clustering of the predictions into vector geometries. An instance of this approach is Weighted DBSCAN, a density-based clustering algorithm that generates parking lot polygons based on pixels, \textit{Map-matching} ~\cite{mapmatching}, which involves matching of predictions to the geo-features already on the network to obtain instance-based predictions, and \textit{filtering} based on custom predicates. Once the post-processing is done, the prediction artifacts are used in downstream pipelines in different ways, viz. for feature detection, anomaly detection, prioritization, etc.

\section{Experiment Lifecycle}
\label{lifecycle}

In this section, we go through the different steps of the experiment lifecycle.
Trinity introduces a simple, intuitive, and declarative approach for tackling problems in the spatial domain through iterative experimentation. The steps involved in solving a problem are problem formulation, label collection, project creation, experiment setup and model training, evaluation, and product deployment.

\begin{figure}[hbt]
	\begin{center}
		\includegraphics[scale=0.20]{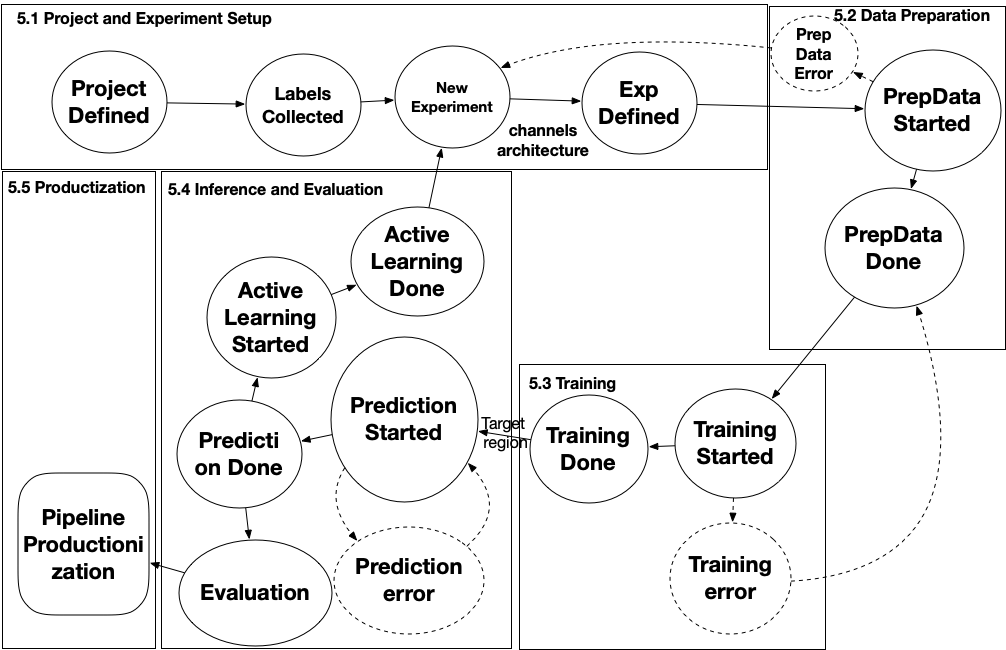}
	\end{center}
	\caption{Project Lifecycle }
	\label{fig:state_diagram}
\end{figure}

Moreover, the states that a single experiment goes through involved in each iteration are explained in Figure ~\ref{fig:state_diagram}.

\subsection{Project and Experiment Setup}

\subsubsection{Project Definition}
A new project space is created to host experiments to test the hypotheses of the users. A project hosts several different experiments to solve a specific problem. 

\subsubsection{Label Collection}
The problem formulation step involves the determination of how to pose the problem and generate labels in a way that is amenable to be solved using Trinity.
In this step, users generate labels for the task. This is done in several ways, either using our in-house label creation tool or by uploading a file containing geometries. A snapshot is provided in Figure ~\ref{fig:lava}.

\subsubsection{New Experiment Definition}
 Experimentation is an iterative process and Trinity makes it really easy to perform several different experiments in parallel with multiple combinations of profiles or hyperparameters. The next step is to define an experiment and that is done by specifying three inputs. They are \textit{Labels} -  which are brought in by the user, \textit{Channels}-selected from a catalog of profiles exposed in the UI, and model architecture that is also selected from a predefined list.

\subsection{Data Preparation}
Once the experiment is set up with the above selections the user needs to just click on a button to launch data preparations jobs on Spark/HDFS for preparing multichannel ``images" for labeled areas and can pre-schedule neural network training on the GPU cluster without needing to know anything that happens underneath.

\subsection{Training}
Once the data is prepared, the training starts. The progress of the training can be monitored using metrics that are generated. If dissatisfied with an experiment, the user can launch more experiments by changing/adding profiles, selecting different neural network architectures or adding more labels. Advanced users like data scientists can also change a few hyperparameters like batch size and learning rates in order to make the training process better and converge faster.

\subsection{Inference and Evaluation}

\subsubsection{Prediction}

Once the model is trained, predicting on a new region is relatively straightforward. The user just uses the Prediction UI to define a prediction by specifying the target region where the predictions are required. Multiple predictions on different areas can be launched from the same experiment simultaneously. The user also has an option to use data of a different time period during inference. For instance, the model could have been trained with some duration of data in one year but the predictions can be done with the same duration of data in a different year and region. Users can also check the results visually as a heat-map, and can post-process them into geometries that can be compared against each other as described in Section \ref{postprocessing}. 
\subsubsection{Evaluation}
All of the above steps can be repeated for several different experiments, testing varied hypotheses with the help of different channels,  label representations, and neural net architectures. Finally, the predictions in validation regions are compared against golden datasets by the users using quantitative and qualitative metrics in order to select the best model for the given use case. We use precision, recall, F1-Score and IOU as metrics for comparison.

\subsubsection{Active learning}
Once the prediction is done, users have the ability to perform active learning to obtain new labels from the users for the next iteration. Once the active learning is done, the new labels selected are added to the label collection for the next iteration of the experiment.

\subsection{Productization}

Once the best model for the use case is identified, the users work on transforming the trained \emph{model} into a \emph{product}. Trinity provides users with a common framework to generate a workflow for a production pipeline that does data pre-processing, prediction, post-processing and also can be integrated with other pipelines, without any overhead of the UI. These product pipelines enable the Trinity models to be packaged neatly so that they can be used in production effectively. Although, not every use case needs downstream pipelines as they may be ad-hoc work for one-time fix or cleanup - in which case predictions made from the UI can be directly used without engineering overheads. 

There are several categories of products that are generated based on the models. 

\textbf{Reference Layer}: One variant of products are \textit{reference layers}, where the desired output is a reference heatmap that aids human editors with their decisions. For instance, when coding the pedestrian crosswalks for a new region, having the reference layer of Trinity predictions is very valuable.

\textbf{Anomaly Detector} Another variant is an anomaly detector that helps detect instances where the existing map has anomalies. Here, given a map, the model makes the prediction and potential errors in the maps are flagged. These errors are then evaluated by humans to fix and improve the map. This is useful for the long-term maintenance of the map for keeping it accurate in a rapidly changing world.

\textbf{Prioritization Filters} In this variant, model output is used to prioritize or filter other signals e.g. certain workloads can be prioritized in commercial areas or complex intersections as detected by a Trinity model. 

\textbf{Evaluator} In this variant, the predictions are used to judge the quality of different data sources to pick the best one from and to evaluate internal and external vendor data.

\section{Sample Applications}

Trinity has been used for a variety of applications using heterogeneous channels many a time in unconventional ways. Some applications are road center-line detection (Figure ~\ref{fig:centerline}) based on driving behavior,  stop signs ( Figure ~\ref{fig:stopsigns}) based on heading profiles, type of road detection, residential areas using temporal embeddings (Figure ~\ref{fig:residential}), building footprint detection based on multiple modalities including satellite imagery, driving footprint, etc. Visualizations of some of these products are included in the appendix ~\ref{sec:appendix_sampleapplications}.

In addition to the above, many more use cases can be addressed using Trinity. Examples include traffic light and stop sign detection, sports area detection, complex intersections, barriers, roundabouts, and missing commercial places.

There are many use-cases already being solved by geospatial experts using the platform and there are many more that can be solved as long as the problem is correctly formulated and suitable labels are collected. We have observed that the use-cases that took weeks to formulate and test out can now be done in hours to days.

\section{Benefits}

\begin{figure}[hbtp!]
	\begin{center}
		\includegraphics[scale=0.28]{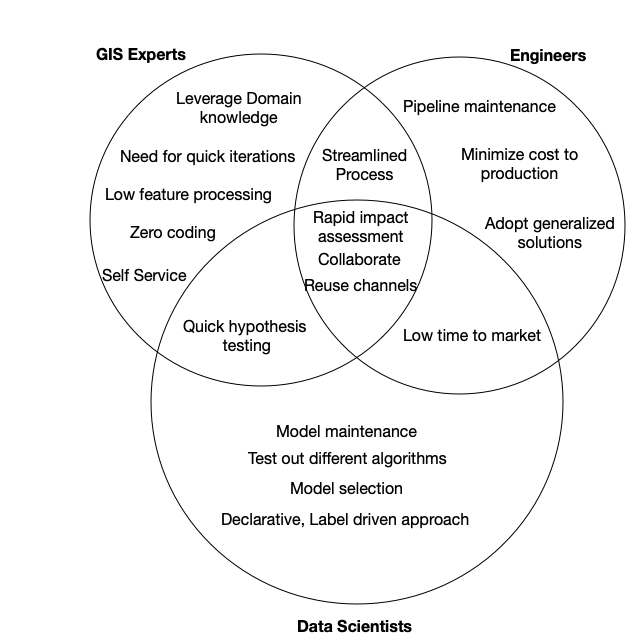}
	\end{center}
	\caption{Categories of users}
	\label{fig:users}
\end{figure}

In this section, we look at the benefits of a platform like Trinity for different users. Three main categories of beneficiaries are domain experts, data scientists, and engineers. These advantages are summarized in Figure ~\ref{fig:users}. % As represented in the Figure \ref{fig:initiated}, we also observe a majority of the new projects now being initiated by the domain experts, something that was not possible earlier.

\subsection{Benefits for the Domain Expert}

\textbf{Enablement:} Trinity enables domain experts to run experiments first hand - a capability that traditionally has been either non-existing or extremely constrained thereby discouraging their direct participation. Trinity makes it possible for the domain experts to make a quick transition from having a promising intuitive idea to rolling their sleeves and building models that validate their hypotheses.

\textbf{Zero Coding:} Trinity is designed as a zero coding platform where the complexities of ML workflow are abstracted out.  It offers a self-service model that enables a wide range of non-technical users to train, predict and scale their own models. This lowers the barrier to entry and domain experts with any level of technical skills are qualified to use the platform.

\textbf{Rapid experimentation:} It also facilitates quick and rapid experimentation through a simple and intuitive UI, backed by powerful GPUs and Hadoop clusters. Rapid experimentation gives domain experts freedom to try out many different ways of solving their problems using their best judgment and iterate based on the results.  

\textbf{Shared Vocabulary and workspace:} Trinity lets the domain experts speak the same language as engineers and data scientists. They participate in discussions on what new channels to create for tapping into more information inherent in various data sources and what combination of channels to use in solving different problems. They are able to ideate, share and collaborate on experiments and predictions with scientists as equal partners. This, in turn, ensures that nothing is lost in translation.

\subsection{Benefits for the Scientist}

	\textbf{Decoupled channel generation:}  Machine learning practitioners work actively in extracting information from the raw data in the form of channels that are later used by segmentation models. This decouples channel generation from downstream products. Scientists find it easy to work on channel generation and scale them independently if needed. Others may choose to use the channels already pre-populated in the channel store.
	
	\textbf{Minimal Feature Processing:} Complex and repeated feature engineering constituted the biggest workload of scientists before. They now reuse and mix \& match channels from other scientists and engineers while building their models. Scientists who are not trained in deep learning also get to leverage advancements in the field directly. 

\textbf{Rapid Experimentation:} The ability to perform rapid and parallel experimentations for a problem helps scientists move and develop faster. Moreover, the ability to compare experiments easily helps them shorten the path to an optimal solution.

\textbf{Enhanced Productivity:} Due to quick and easy experimentation and minimal feature processing, we have observed that workflows that took weeks or months to set up and train can now be tested out in a matter of hours and days. There's low latency from idea to initial prototype.  Standardized problem formulation, workflows, and reusable channels lead to enhanced productivity.  With the tools to automate much of their boilerplate code, data scientists can focus more on the solutions to business problems.

\textbf{Better Collaboration:} As discussed above, having a shared vocabulary in terms of channels, models and labels facilitates better collaboration between team members working on related projects in Trinity. Experiments are often shared or passed over between scientists and domain experts seamlessly and communication gets easier.

	\textbf{Scalablilty:} Trinity is designed to be horizontally scalable during both training and inference. This alleviates the need for the scientist to be dependent on the engineer to scale up the jobs before they can be tested at scale. 

\textbf{Go to market:} Trinity provides a standardized workflow to the scientists and lays out a clear path to the productionization of the model.

\subsection{Benefits for the engineer}

\textbf{Modularity:} Trinity allows engineers to be decoupled from the problem statement and model building. Engineers contribute in pluggable modules ranging from scaling channel creation jobs, building the platform components all the way to writing generic data processors. 

\textbf{Streamlining:} Trinity enables engineers to avoid duplication of work and easier maintenance by promoting standardized workflows for diverse problems. This standardized training and inference take away the need for performing repetitive customized tasks for different use cases and flavors of machine learning. This consolidation ensures that any effort spent in optimizing workflows now benefits all the users and use cases. This also helps enable an engineer to maintain and debug any downstream product with the least amount of knowledge transfer.

Additionally, given the intuitive nature of problem-solving, other non-expert stakeholders and higher management tend to understand the solution to critical problems better and participate more actively. They tend to have higher confidence in the planned efforts as there is a clear path from problem statements to a deployed solution.

\section{Future Work}

Trinity paves the way for some high-impact innovations in the geospatial domain.  We intend to expand the capabilities of the platform by investing in the following techniques in near future: 

\textbf{Data and model governance} Although versioning of artifacts such as models and the channels are already supported by the system, we hope to make the versioning and provenance tracking a first-class citizen of the platform. 

\textbf{Drift monitoring}: Drift monitoring refers to the tracking of data drifts in the input channels and concept drifts in the applications so that the staleness of models is caught and alerts are raised as soon as possible.

\textbf{Meta Learning}: Meta-learning \cite{meta1} can be used to improve related approaches and capabilities already available in Trinity like transfer learning \cite{meta2}, and multi-task learning \cite{meta6}. Specifically, for a new target task, the model repository of trained models can be used as a source of meta-knowledge to provide an estimate of the initial model parameters of the neural network \cite{meta2}, the neural architecture, or the optimization strategy. We believe that combined with NAS and hyperparameter optimization, meta-learning will enable users to focus more on the problem by abstracting away the complexities involved in optimal parameter and configuration selection. Recent advances reported in literature (\cite{meta7}, \cite{meta8}) also demonstrate that meta-learning can benefit neural architecture search, domain adaptation and generalization, and continual learning.

\textbf{Uber Model}: Currently, Trinity requires users to select and experiments with different sets of channels for every problem independently. This limits the knowledge sharing between tasks. Multitask support in Trinity helps cross-task learning to some extent but we believe that having a big single model or Uber model that solves most of the seemingly different tasks together will help in generalizing better. The single model will also take away the need of guessing the combination of channels since all of them will be used by the model, thereby giving the opportunity to use dense channels like embedding fields more actively. We hope that the Uber model backed by AutoML and NAS will not just help find better solutions to the problems but also make it easy for domain experts who can then just focus on problem formulation.

\section{Discussion and Conclusion}
This work presents the motivation, architecture, and working of Trinity,  a code-free, self-service platform that brings the power of deep learning for spatial datasets including imagery to a diverse set of use cases in the geospatial domain. It enables the users to harness the power of semantic segmentation on spatio-temporal datasets. It promotes quick and easy experimentation, improves productivity, collaboration and standardizes the machine learning workflows thereby minimizing duplication of effort. It lowers the barrier to entry for non-technical domain experts who generally are the best people to solve problems of the respective domain. 

There are many tangible and intangible benefits of ``No-Code" domain-specific AI platforms. Even though Trinity is custom-built for a specific domain, we believe that this platform can be as easily deployed for any domain where the input data is already like or can be converted to image-like channels and problems formulated as a computer vision problem. Furthermore, it lays out a template to build wide and deep platforms for other domains by narrowing down on suitable  AI techniques after appropriate signal transformation and intelligent problem formulation. Real \textit{democratization} of AI happens when domain experts are enabled to tap into the raw complex signals of their domain and solve intricate problems on their own with almost everything else including data, information, knowledge extraction, algorithms, software frameworks, hardware dependencies, and platforms abstracted out by engineers and scientists. We believe that Trinity is a step towards such a future.

\bibliographystyle{plain}

\begin{appendices}

\clearpage
\section{Trinity Applications}
\label{sec:appendix_sampleapplications}

	\begin{figure}[!hpb]	
	\begin{center}
		\includegraphics[scale=0.20]{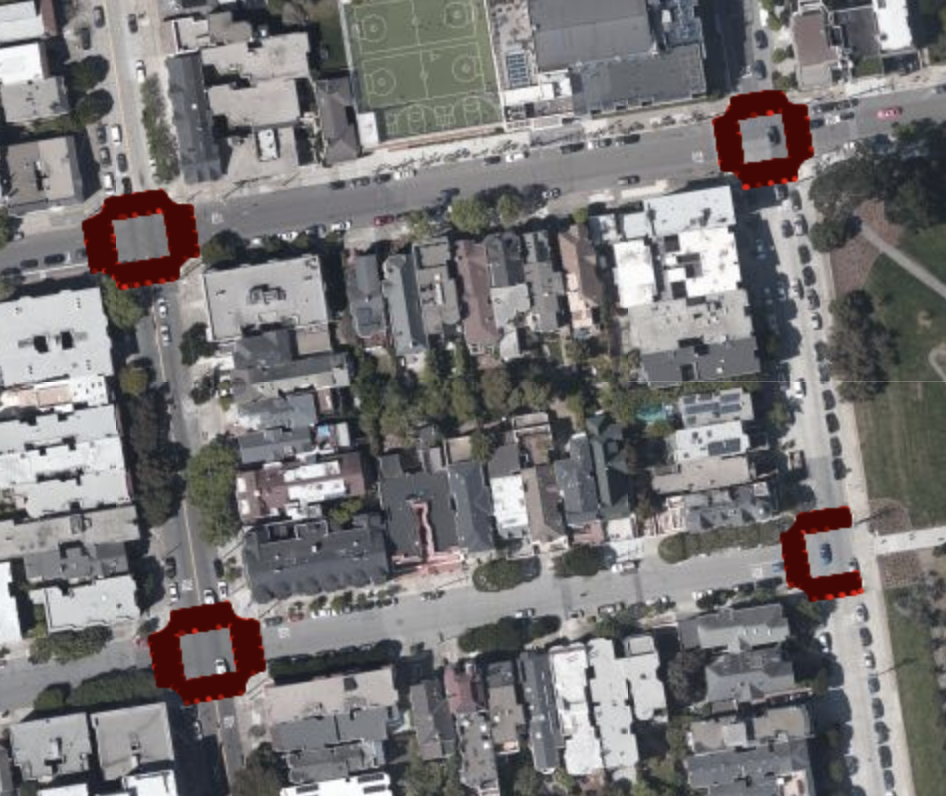}
	\end{center}
	\caption{Binary segmentation: Detection of pedestrian crosswalks with imagery and motion based channels}
	\label{fig:binarysegmentation}
	\end{figure}

\begin{figure}[!hpb]
	\begin{center}
		\includegraphics[scale=0.15]{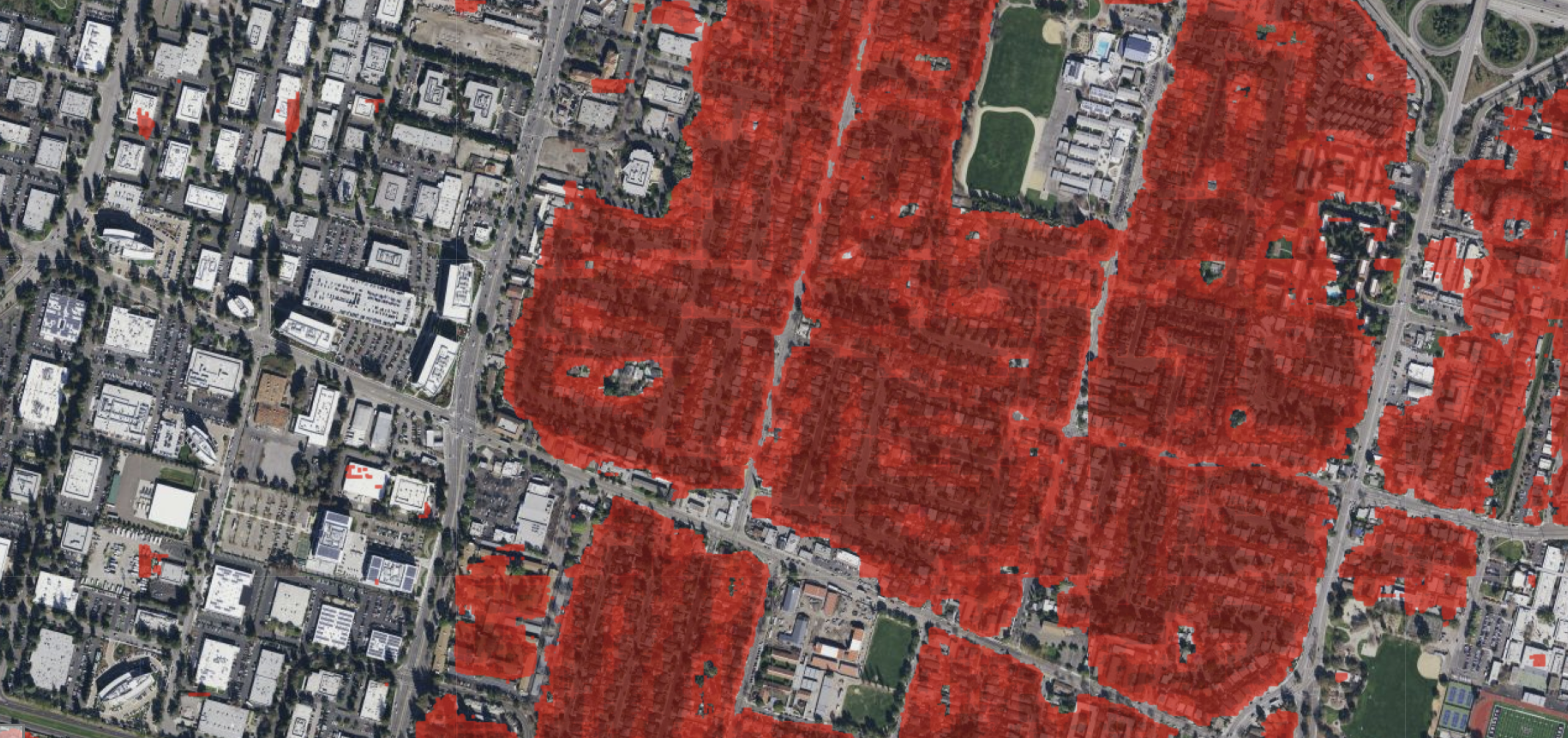}
	\end{center}
	\caption{Residential Area Detection motion based channels (imagery only for reference)}
	\label{fig:residential}
\end{figure}

\begin{figure}[!hpb]
	\begin{center}
		\includegraphics[scale=0.20]{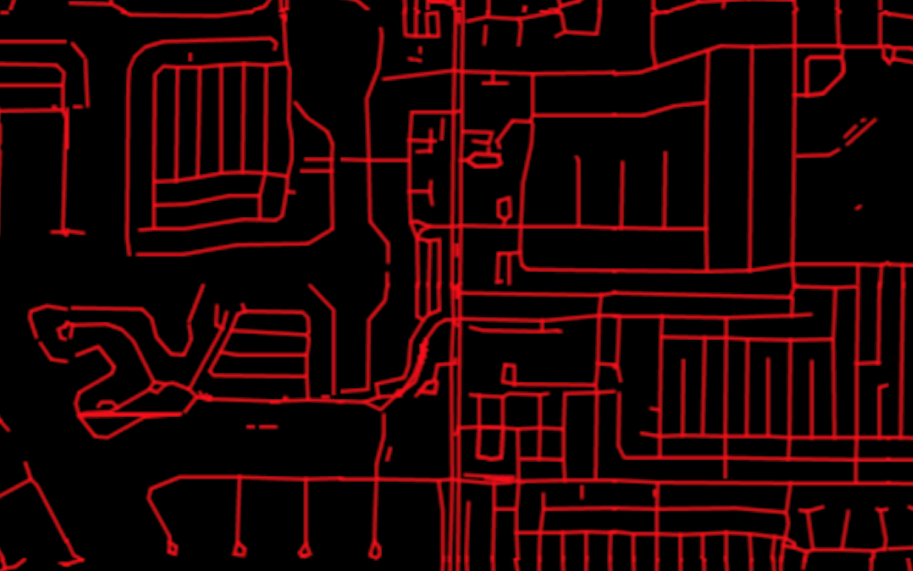}
	\end{center}
	\caption{Predicted road centerlines using motion based channels}
	\label{fig:centerline}
\end{figure}

\begin{figure}[!hpb]
	\begin{center}
		\includegraphics[scale=0.24]{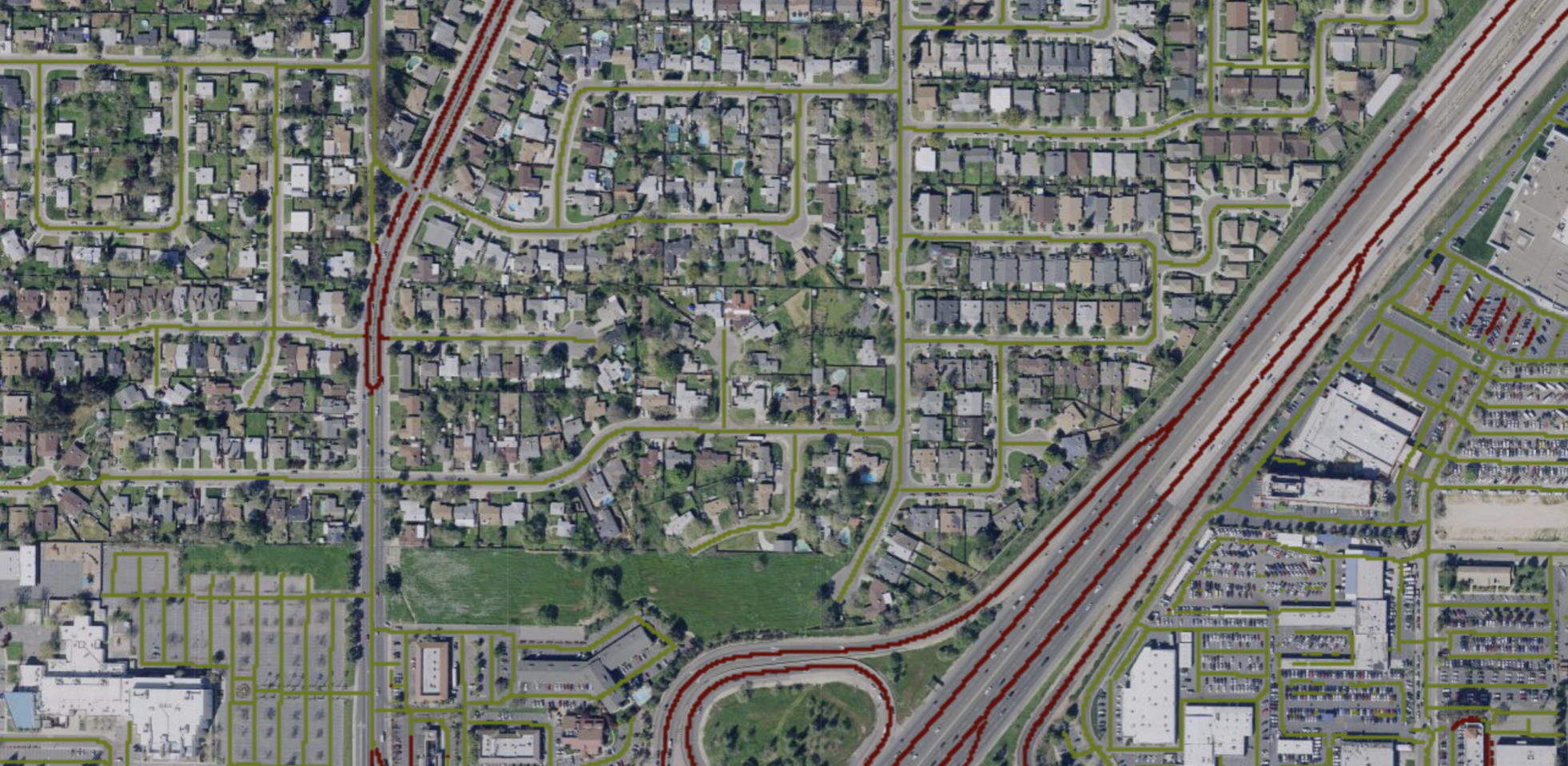}
	\end{center}
	\caption{Oneway detection using motion based channels (imagery only for reference)}
	\label{fig:oneway}
\end{figure}

\begin{figure}[!hpb]
	\begin{center}
		\includegraphics[scale=0.12]{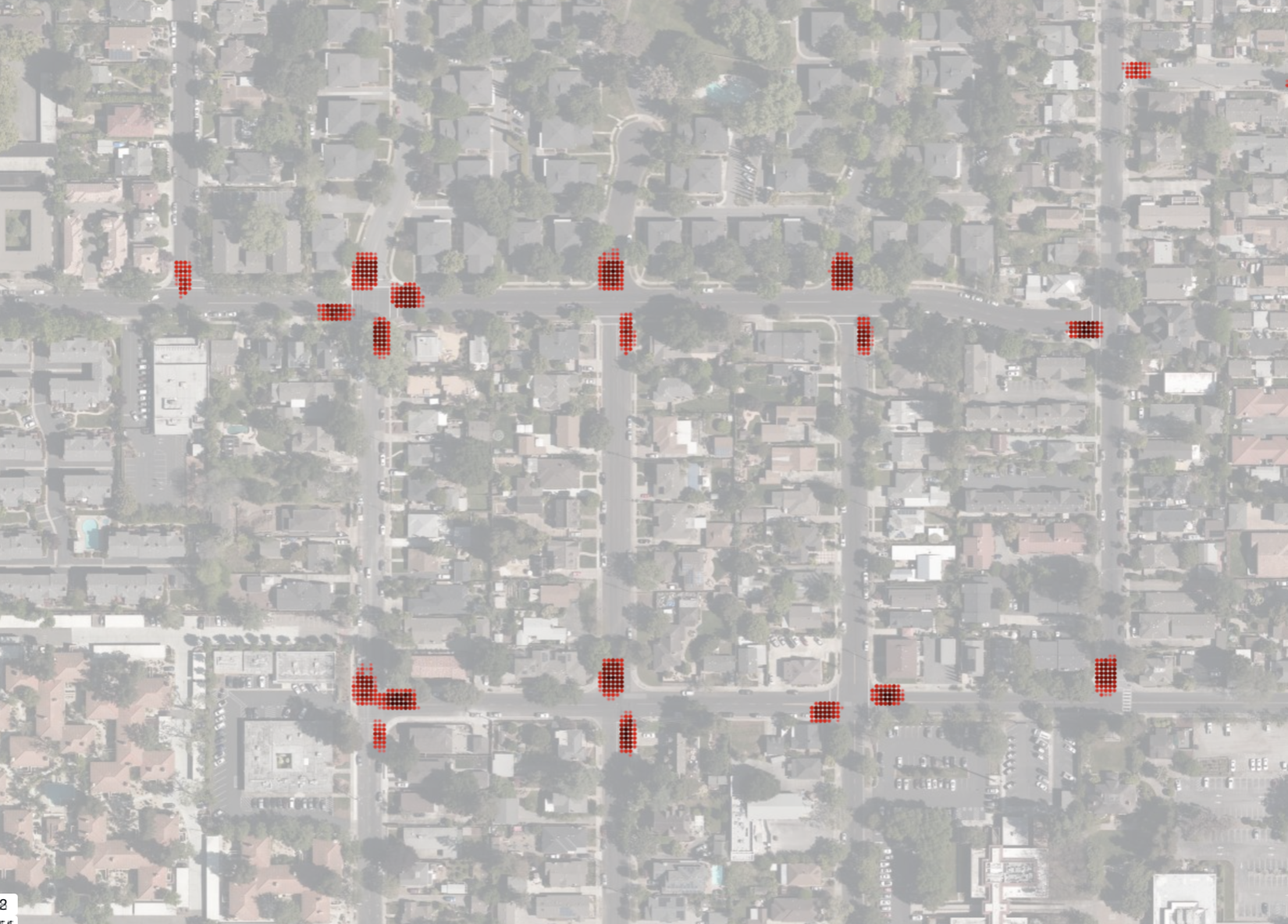}
	\end{center}
	\caption{Predicted stop sign locations using motion based channels (imagery only for reference)}
	\label{fig:stopsigns}
\end{figure}

\begin{figure}[!hpb]
	\begin{center}
		\includegraphics[scale=0.20]{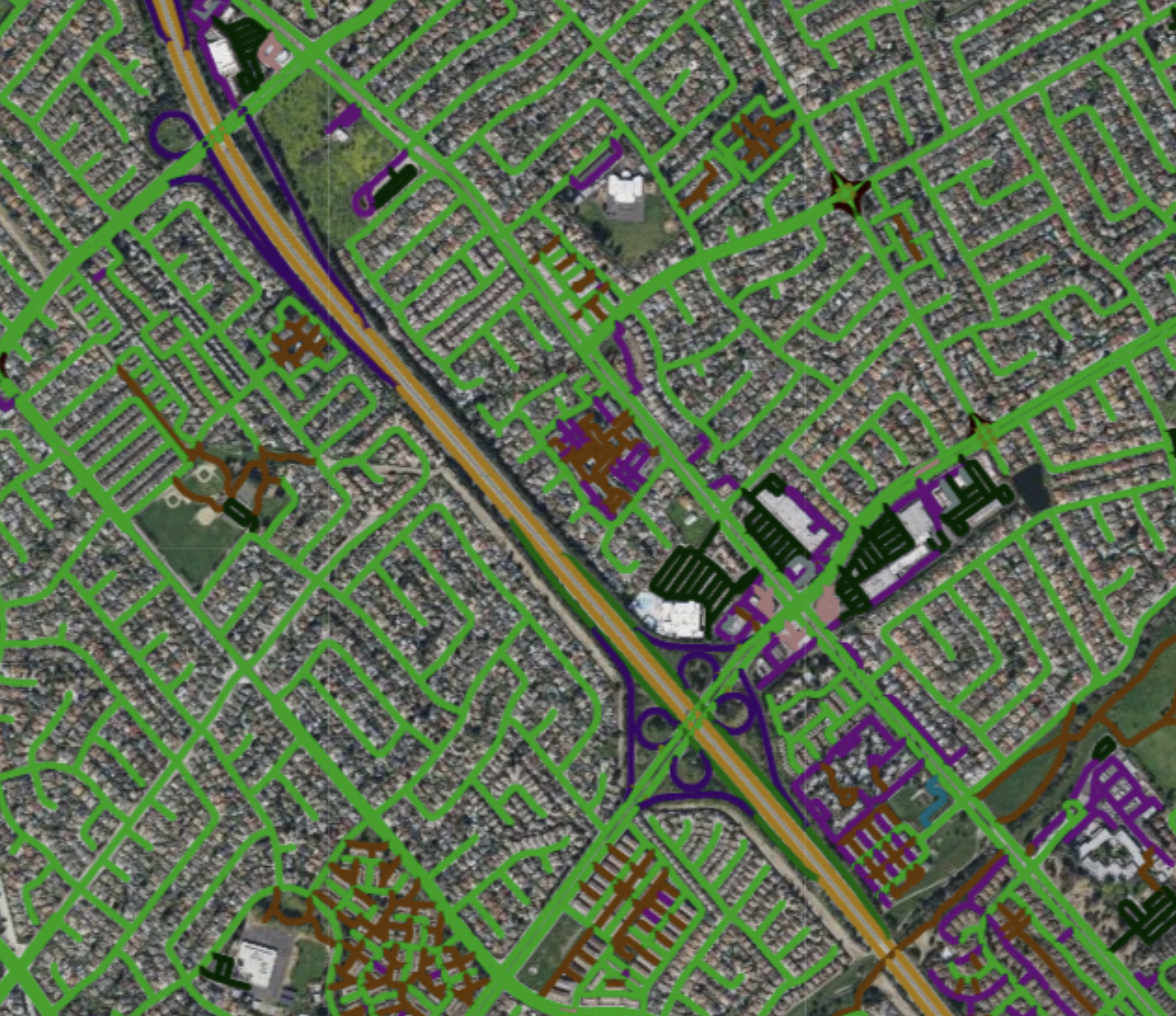}
	\end{center}
	\caption{Multiclass Type of Road detection using motion based channels (imagery only for reference)}
	\label{fig:fow}
\end{figure}
\end{appendices}

\end{document}